\documentclass{sig-alternate}
\usepackage{graphicx}
\usepackage{balance}  
\usepackage{url}
\usepackage[colorinlistoftodos]{todonotes}
\newcommand{\richard}[1]{{#1}} 

\author{\alignauthor Richard McCreadie, Craig Macdonald and Iadh Ounis  \\
\email{\{firstname.lastname\}@glasgow.ac.uk}\\
\affaddr{University of Glasgow, UK}\\
}

\conferenceinfo{SWDM 2016}{October 28th, Indianapolis, USA}
\copyrightetc{}

\begin{document}

\title{Emergency Identification and Analysis with EAIMS}

\date{6 October 2016}

\maketitle

\begin{abstract}
\looseness -1 Social media platforms are now a key source of information for a large segment of the public. As such, these platforms have a great potential as a means to provide real-time information to emergency management agencies. Moreover, during an emergency, these agencies are very interested in social media as a means to find public-driven response efforts, as well as to track how their handling of that emergency is being perceived.  However, there is currently a lack advanced tools designed for monitoring social media during emergencies. The Emergency Analysis Identification and Management System (EAIMS) is a prototype service that aims to fill this technology gap by providing richer analytic and exploration tools than current solutions. In particular, EAIMS provides real-time detection of emergency events, related information finding, information access and credibility analysis tools for use over social media during emergencies. 
\end{abstract}

\section{Introduction}
\richard{Social media platforms are some of the most popular information sharing mediums in use today. They provide a wealth of real-time information, which, if used effectively can provide insights about what is happening in the world. Importantly, during emergency situations such as natural disasters, the information shared on social media has been shown to be valuable to enhance the emergency response~\cite{mccreadie2015SUPER}. In particular, experts in a variety of domains have developed tools to use social media to identify and track events~\cite{Abel:2012:TFF:2187980.2188035,heim2011semsor,imran2014aidr,osborne2014,purohit2013twitris,rogstadius2013crisistracker}. For example, the Crisis Tracker system~\cite{rogstadius2013crisistracker} makes use of a manually created set of search terms to crawl crisis-related content, which is then clustered automatically and manually annotated by volunteers. In general, social media tracking and analysis tools can collect, enrich, aggregate and visualize information from people involved in an event, enabling both the general public and emergency response agencies to better monitor what is happening in real-time.}
\enlargethispage{1\baselineskip}

\richard{However, many current crisis tracking and analysis tools rely on real-time volunteer (``the crowd'') effort to enable or enhance core functions, such as information categorization~\cite{imran2014aidr}, location identification~\cite{okolloh2009ushahidi} or information verification~\cite{rogstadius2013crisistracker}. This reliance on humans for core functionalities can be disadvantageous, particularly during the outset of a disaster, where volunteers may not yet be in place. Furthermore, current fully automatic platforms such as Twitris~\cite{purohit2013twitris} and Twitcident~\cite{Abel:2012:TFF:2187980.2188035} are tied to one social media platform, and hence are not applicable to deployment in regions where those platforms are less prevalent. On the other hand, a wider array of automatic solutions for real-time event detection in social streams~\cite{mccreadie2013sigirEDStorm}, sentiment analysis~\cite{ICWSM16vargas}, timeline summarization~\cite{McCreadie:2014:IUS:2661829.2661951} and credibility estimation of social content are becoming available, which present new opportunities to provide enhanced automatic tools for disaster monitoring on social media.}

\richard{Hence, we present a new crisis tracking and analysis toolkit, named: EAIMS (Emergency Analysis Identification and Management System), which aims to provide automatic identification, tracking, summarization and exploration tools over social content generated during emergency events. EAIMS differs from most existing crisis analysis toolkits, in that it provides a richer suite of automatic tracking and analysis tools, namely: automatic event detection, timeline construction, targeted sentiment analysis, user-community identification and information credibility estimation, as well as information access support tools in the form of social media search and discussion threads ranking and visualization. EAIMS is social media platform-independent, making it applicable to deployment in regions that use alternative platforms.}

\begin{figure*}[t]
\centering
\includegraphics[width=175mm]{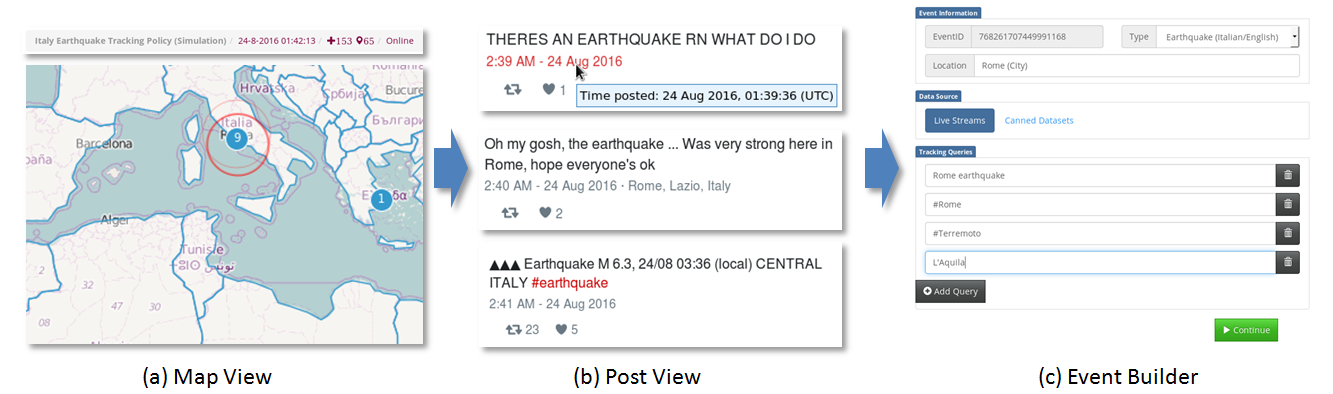}
\caption{Strategic Layer Views.}\label{fig:stratV}
\vspace{-3mm}
\end{figure*}

\begin{figure*}
\centering
\includegraphics[width=175mm]{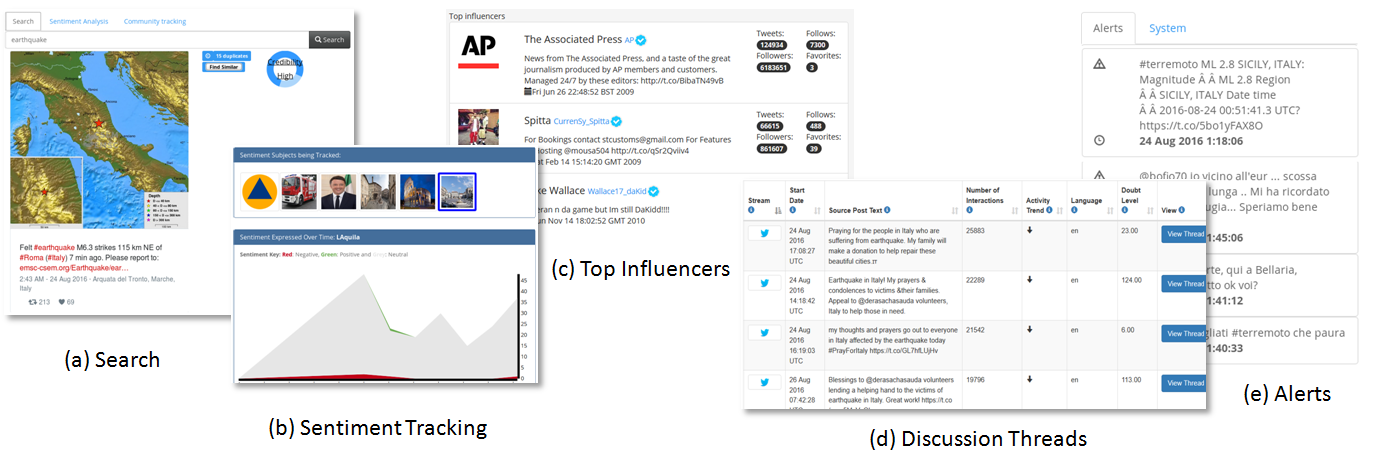}
\vspace{-2mm}
\caption{Tactical Layer Views.}\label{fig:tactV}
\vspace{-3mm}
\end{figure*}
\section{EAIMS Functionality}
\richard{To illustrate the functionalities of the EAIMS toolkit, we describe how EAIMS might be used during a real event. More precisely, we use the large earthquake that occurred during 24th of August 2016 in central Italy as our example when using Twitter as a social media data source, as illustrated in Figures~\ref{fig:stratV} and \ref{fig:tactV}. }

\richard{EAIMS supports two operational levels, namely strategic and tactical, based on requirements defined by civil protection and law enforcement agencies~\cite{mccreadie2015SUPER}. The strategic level supports the early identification of an event for the end-user using social media as a sensor for what is happening in the world. In particular, users are able to define tracking profiles targeted for particular types of events (earthquakes in our example). Once a profile is registered, social media content is collected and first story detection (FSD) technologies~\cite{mccreadie2013sigirEDStorm} are used to identify posts representing candidate events, which are then pushed in real-time to a map screen as pings (Figure~\ref{fig:stratV} (a)). The user can view the content of these posts within the strategic UI (Figure~\ref{fig:stratV} (b)). Multiple tracking policies can be used concurrently by different users, which enables the identification multiple events of different types. Once an event has occurred, the user can transition from the strategic to the tactical level. The tactical level provides a targeted data collection for a particular event, as well as a suite of data enrichment, data exploration and visualization tools for use of the collected content. In particular, once the user is confident that an emergency event is in progress, they can use the \emph{Event Builder} function of EAIMS to configure the tracking of that event, e.g. by setting event-specific tracking terms to collect data for (Figure~\ref{fig:stratV} (c)). Once the event tracking is configured, the user opens the tactical level UI for that event. The tactical level UI visualizes the output of a series of social sensors that select or enrich (e.g. by automatically adding sentiment or credibility labels) the social data collected, as well as provides social media content search and discussion threads visualization. In particular, within the tactical UI, EAIMS provides: real-time search using the Terrier Information Retrieval platform~\cite{macdonald2012puppy} with automatic credibility estimation over the ranked content (Figure~\ref{fig:tactV} (a)); sentiment tracking for particular named people, organizations or places (Figure~\ref{fig:tactV} (b)); top influencer identification within the social network(s) (Figure~\ref{fig:tactV} (c)); visualization of the most active relevant discussion threads within the social network(s) (Figure~\ref{fig:tactV} (d)); and automatic event timeline generation (Figure~\ref{fig:tactV} (e)). The underlying EAIMS services that feed the tactical visualization support content collection and analysis from multiple events concurrently, as long as those events have been registered with the event builder.}

\section{Recent Additions}
An earlier version of the EAIMS toolkit was published in \cite{EAIMS}, where the additional technical details of the platform can be found. The current version of the toolkit has been extended with real-time automatic extraction and storage of active discussion threads within the social media platform, as well as ranking and visualization capabilities for those discussion threads (Figure~\ref{fig:tactV} (d)). 

In particular, based on consultation with civil protection and police agencies, it was suggested that discussion threads on social media are a desirable means to both represent discussion topics and user communities during emergencies. Indeed, when considering reasoning about information veracity within social networks, discussion threads are a natural unit to analyze, as they are topically coherent. 

However, within some social networks, discussion threads cannot be accessed directly. Instead, discussion threads are exposed as annotations on each individual post. For instance, within Twitter, discussion threads are represented by either `in\_reply\_to\_status\_id' or `retweet.id' annotations added to each tweet. For this reason, it is necessary to construct the discussion thread structure in real-time. 

EAIMS uses an inverted index structure to construct the various discussion threads for an event over time. In this case, the inverted index stores the posting list (documents contained) for each discussion thread, while the lexicon stores the identifiers for each thread. When a new post $t_n$ arrives on the stream, if that post is not a reply to a previous post, then it is considered the start of a new thread, and is hence added to the lexicon (we use the identifier of the first post as the thread identifier). If the post $t_n$ is a reply, then a three-stage update operation is performed. First, the inverted index is used to get the posting list for the source post that tn replies to, denoted $t_s$. This posting list is in effect the discussion thread that $t_n$ belongs to. Second, a new `document' is generated that contains the identifiers from all of the posts in the $t_s$ posting list, as well as the identifier for $t_s$, i.e. this `document' represents the new thread for $t_n$. This document is indexed, which both constructs the posting list for $t_n$, but also updates the posting lists for the (sub-)threads within $t_n$. 

Conceptionally, under this approach, the inverted index maintains k copies of a discussion thread, where k is the length of that thread. Each copy is keyed by a different post within the thread. In this way, the full discussion thread can be retrieved in a single operation regardless of which post in the thread is replied to. For long-running events with very large thread sizes, storing multiple copies of each thread can be avoided by having all k posts within a thread `map' down to a single posting list.

\begin{figure}[t]
\centering
\includegraphics[width=84mm]{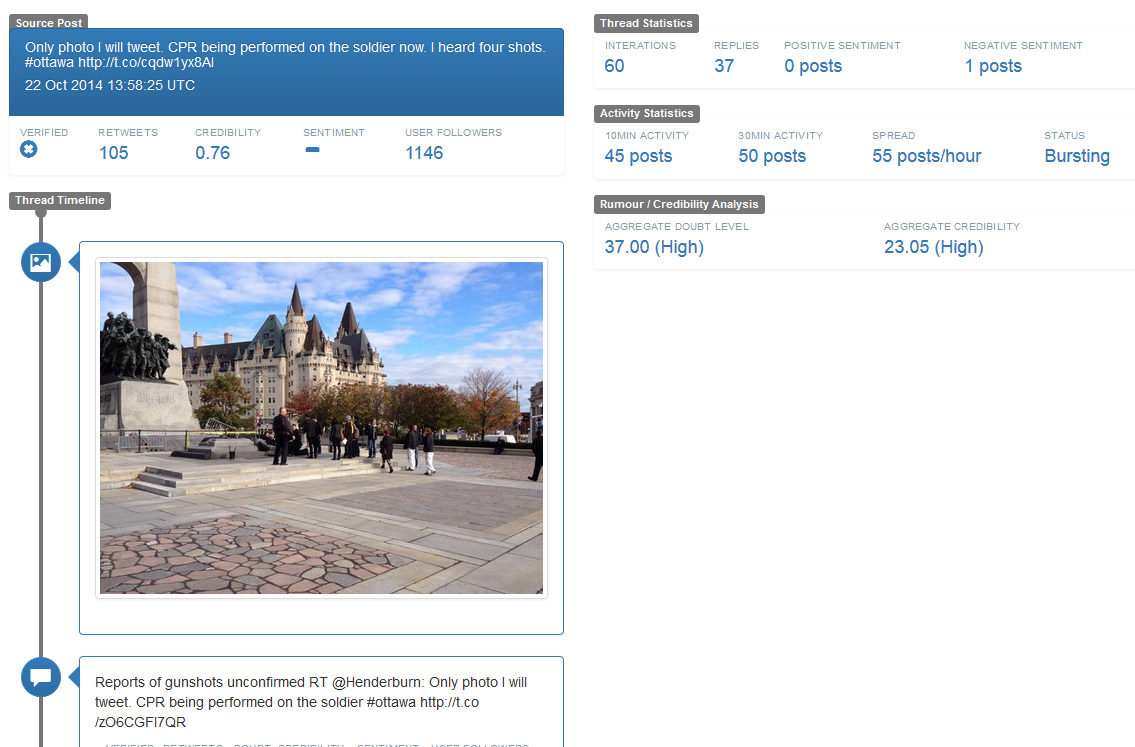}
\caption{Discussion Thread Visualization.}\label{fig:thread}
\end{figure}

Within EAIMS, the inverted index structure can be used to find the largest discussion threads at the current moment for the user. These are visualized as a tabular overview, as shown in Figure~\ref{fig:tactV} (b). A user can then select a particular thread to open a more detailed view of the post within that thread and their statistics, as illustrated in Figure~\ref{fig:thread}.

\section{Outlook}
EAIMS is a prototype tool designed to aid emergency management organizations analyze and explore social media content in real-time during emergency and/or large public events. It was developed as part of a larger EC FP7 project called SUPER (Social sensors for security assessments and proactive emergencies management). The outcomes of SUPER, including the social sensors illustrated by EAIMS are being deployed and evaluated by end-users from the Civil Protection Service of the Campania Region (Italy) during flash flood emergencies, as well as end-users from the General Inspectorate of Romanian Police, for safety monitoring during large public events.   

\balance

\enlargethispage{1\baselineskip}

\section{Acknowledgments}
This work has been carried out in the scope of the EC co-funded SUPER Project (FP7-606853).

\bibliographystyle{abbrv}

\end{document}